\documentclass[prl,twocolumn,showpacs,aps,superscriptaddress]{revtex4-1}
 
\usepackage{amsmath}
\usepackage{amssymb}
\usepackage{amsmath}
\usepackage{txfonts}
\usepackage{graphicx}
\usepackage{epsfig}
\usepackage[T1]{fontenc}
\usepackage[latin9]{inputenc}
\usepackage{times}
\usepackage{color} 
\usepackage{xspace}
\usepackage{amssymb,amsmath}
\usepackage{amsbsy}
\usepackage{braket}
\usepackage{tabularx}
\usepackage{mathtools}          %loads amsmath as well
\usepackage{capt-of} 
\DeclareMathOperator{\sech}{sech}

\usepackage{ifpdf}
\ifpdf
\usepackage{epstopdf}
\fi

\usepackage[unicode]{hyperref}
\hypersetup{
	pdftitle={spinnorBEC},
	pdfauthor={Yu},
	pdfsubject={Vortexlines},
	colorlinks=true,
	linkcolor=blue,
	citecolor=blue,
	filecolor=black,
	urlcolor=blue
}
\def\urlprefix{}
\def\url#1{}

\usepackage{bm}% bold math
\usepackage{color}
\newcommand{\be}{\begin{equation}}

\newcommand{\ee}{\end{equation}}

\newcommand{\bea}{\begin{eqnarray}}

\newcommand{\eea}{\end{eqnarray}}

\newcommand{\nn}{\nonumber }

\begin{document}

%\title{Two-dimensional dynamics of a ferrodark soltion by seeded snake instability:  Absence of decay and proliferation of confined vortex dipoles } 
%\title{Dynamics of ferrodark soliton rings}

%\title{Oscillating ring ferrodark solitons and nematic breather modes at the core in a homogeneous spinnor superfluid}

%\title{Oscillations of ring ferrodark solitons and nematic modes at the core in a homogeneous superfluid}

% \title{Dynamics of ring ferrodark solitons and nematic breather modes at the core}
% \title{Oscillating ring ferrodark solitons and the ring tension induced nematic modes at the core}

%\title{Ring ferrodark soliton oscillations and nematic breathing modes at the core  in a  homogeneous spinor superfluid }
 
\title{Oscillating ring ferrodark solitons with breathing nematic core in a homogeneous spinor superfluid}

%\title{Ring ferrodark solitons in a homogeneous spinor superfluid: radius oscillation and nematic breathing modes at the core}

\author{Xiaoquan Yu}
\email{xqyu@gscaep.ac.cn}
\affiliation{Graduate School of China Academy of Engineering Physics, Beijing 100193, China}
\affiliation{Department of Physics and Centre for Quantum Science, University of Otago, Dunedin, New Zealand}
%
%\author{P.~B.~Blakie}
%\affiliation{Department of Physics, Centre for Quantum Science, and Dodd-Walls Centre for Photonic and Quantum Technologies, University of Otago, Dunedin, New Zealand}

\begin{abstract} 
We study the dynamics of ring ferrodark solitons (FDSs) in a homogeneous quasi-two-dimensional (2D) ferromagnetic spin-1 Bose-Einstein condensate (BEC).  In contrast to the usual expanding dynamics of ring dark solitons in a homogeneous system,  the ring FDS radius exhibits self-sustained oscillations accompanied by the nematic tensor breathing at the magnetization-vanishing ring FDS core.  When the ring radius greatly exceeds the FDS width, motion is nearly elastic, and we derive the ring-radius equation of motion (EOM) which admits exact solutions. This equation can be recast into a form analogous to the inviscid Rayleigh-Plesset equation governing spherical bubble dynamics in classical fluids, but with anomalous terms.
At the ring FDS core,  the nematic tensor motion is parameterized by a single parameter that connects the two types of FDSs monotonically.  Beyond the hydrodynamics regime, density and spin wave emissions become significant and cause  energy loss, shrinking the ring FDS radius oscillation; below a threshold, collapses occur followed by the ring FDS annihilation. 
In the zero quadratic Zeeman energy limit,  the ring radius and  eigenvalues of the nematic tensor become stationary,  while oscillations of the nematic tensor components, driven by the ring curvature,  persist at the core.   Excellent agreements are found between analytical predictions and numerical simulations.  

\end{abstract}

\maketitle

\textit{Introduction---}
Dynamics of topological solitons (kinks and domain walls) is rich and is of interests in many research areas ranging across fields from condensed matter~\cite{chaikin1995principles, nelson2002defects} to cosmology~\cite{vachaspati2007kinks}.  In superfluids, much attention has  been devoted into their dynamics in quasi-one dimensional (1D) systems~\cite{pitaevskii2016bose, Anglin2000, Pitaevskii2004, Pitaevskiisnake2008, Pitaevskii2011, Liao2011, Peng2016, Mateo2022, MDQu2016, Jieliu, FDSexact2022, FDSspincorrection, DSEXP,Collision_BEC2,MSexp1,MSexp2,rabec2025bloch}.  Solitons of circular symmetry or ring solitons might be the simplest configuration introducing 2D features~\cite{snakeinstability}.  Ring dark solitons have been firstly  investigated  in the context of nonlinear optics~\cite{ringdarksolitonKivshar,kivshar1998dark} and later in superfluids~\cite{Ringdarksoliton2003, Xue2004, Theocharis2005,Carr2006, Hu2009,Toikka2013, Wang2015,Kevrekidis2017,He2019,Hikaru2023}.  In a uniform scalar BEC, since the force acting on a ring element caused by the ring tension points inwards and its inertial mass is negative~\cite{Pitaevskiisnake2008,Pitaevskii2011},  hence the ring radius,  starting with zero initial velocity,  grows indefinitely and the expanding velocity approaches the speed of sound asymptotically.  The inertial mass of relevant solitons being negative is general~\cite{Pitaevskii2011,Gallemi2019, Sophie, MDQu2016}, therefore  this scenario revealed by  the scalar ring dark soliton dynamics is rather typical.  In the presence of a harmonic trap, the combination of the  buoyancy  force and the ring tension yields to short-time lived oscillatory motion of a ring dark soliton~\cite{Ringdarksoliton2003,Theocharis2005}.  A natural question is whether there exists a topological ring soliton exhibiting self-maintained and long-lived oscillations \textit{in the absence} of an external potential. To the best of our knowledge,  no such kind of motion has been reported. 
Recently discovered ferrodark solitons (FDSs), manifesting as kinks in the transverse magnetization $F_{\perp}\equiv F_x+i F_y$ (the magnetic field is along the $z$-axis) in ferromagnetic spin-1 BECs~\cite{MDWYuBlair, FDSexact2022, FDScore2022, FDSsnake2024}, could be the candidates due to their special property of  sign change of the inertial mass at the maximum propagating speed.

In this Letter,  we investigate  the dynamics of a ring FDS in the easy-plane phase of  a homogeneous quasi-2D spin-1 BEC. We find that, distinct from the usual scenario of expanding dynamics,  the ring FDS exhibits oscillations  caused by the sign change of the ring FDS inertial mass density~\cite{FDSexact2022}.  In the hydrodynamics regime, i.e., when the ring radius is much larger than the FDS width, the motion is near elastic,  allowing us to derive the ring radius EOM to which exact solutions are obtained. At the magnetization-vanishing ring FDS core, the non-zero eigenvalue of the nematic tensor, i.e., the mass superfluid  density oscillates with the same frequency. 
We find that the dynamics of the  nematic tensor at the ring core can be parameterized by a single parameter which unifies two types of FDSs.  
%The ring radius oscillation period decreases as decreasing the quadratic Zeeman energy.  However, in this limit the analytically predicted ring radius oscillation period does not vanish but saturates to a finite value depending on the ring radius. 
Radiation becomes considerable for small FDS rings,  causing the ring energy dissipation and the oscillation amplitude shrinking. When the radius becomes smaller than a threshold,  the ring FDS collapses and finally annihilates.  An analytical estimation of this threshold is given.    
At zero quadratic Zeeman energy, the ring radius dynamics freezes due to the magnetization conservation and the eigenvalues of the nematic tensor at the core become static, yet the ring curvature drives persistent  oscillations of the nematic tensor components at the core. 

\textit{Spin-1 BECs and FDSs---}
A spin-1 BEC  consists of atoms with three hyperfine states $\ket{F=1,m=+1,0,-1}$~\cite{Ho98,OM98,Stampernatrue2006} and is described by  the three component wavefunction $\psi=(\psi_{+1},\psi_{0},\psi_{-1})^{T}$. The Hamiltonian density of a weakly interacting spin-1 BEC reads 
 \bea 
\label{Hamioltonian}
	{\cal H}=  \frac{\hbar^2 \left|\nabla \psi\right|^2 }{2M} +\frac{g_n}{2} |\psi^{\dag}\psi|^2+\frac{g_s}{2} |\psi^{\dag} \mathbf{S} \psi|^2 +q \psi^{\dag} S^2_z \psi,
\eea
where $M$ is the atomic mass, $g_n>0$ is the density interaction strength, $g_s$ is the spin-dependent interaction strength,
$\mathbf{S}=(S_x,S_y,S_z)$, and $S_{j=x,y,z}$ are the spin-1 matrices~\cite{StamperRMP,KAWAGUCHI12}. The uniform magnetic field is  along the $z$-axis, and  $q$ denotes the quadratic Zeeman energy.  The dynamics of the field $\psi$ is governed by the spin-1 Gross-Pitaevskii equations (GPEs) 
\begin{subequations}
\bea
\hspace{-5mm} i\hbar \frac{\partial \psi_{\pm 1}}{\partial t}&=&\left[H_0+g_s\left(n_0+n_{\pm 1}-n_{\mp 1}\right)+q \right]\psi_{\pm 1}+g_s \psi^2_0 \psi^{*}_{\mp 1},\,\, \\
i\hbar \frac{\partial \psi_0}{\partial t}&=&\left[H_0 +g_s\left(n_{+1}+n_{-1}\right) \right]\psi_0 + 2g_s \psi^{*}_0\psi_{+1}\psi_{-1},
\eea 
\label{spin-1GPE}
\end{subequations}
\hspace{-1mm}where $H_0=-\hbar^2\nabla^2/2M +g_n n$, the component number  density $n_m=|\psi_m|^2$ and the total number density (mass superfluid density) $n=\sum n_m$.  The magnetization density $\mathbf{F}\equiv\psi^{\dag} \mathbf{S} \psi$ serves as  the order parameter quantifying the ferromagnetic order $|\mathbf{F}|>0$ for $g_s<0$ ($^{87}$Rb,$^7$Li)~\cite{Ho98,OM98,Stampernatrue2006,StamperRMP,KAWAGUCHI12}.   
For $0<\tilde{q}\equiv-q/(2g_sn_b)<1$, the easy-plane phase, characterized by $F_{\perp}\neq0$ and $F_z=0$,  is the uniform ground state of ferromagnetic ($g_s<0$) spin-1 BECs~\cite{StamperRMP,KAWAGUCHI12}, where $n_b$ is the ground state total number density.  A FDS is an Ising type $Z_2$ kink/domain wall in the spin order and connects oppositely magnetized domains~\cite{MDWYuBlair, FDSexact2022, FDScore2022}.  There exists two types of FDSs and type-I (II) FDSs have positive (negative) inertial mass.  At the speed limit the type transition (the transition between a type-I FDS and a type-II FDS) occurs~\cite{FDSexact2022} and this property plays a key role in the ring FDS dynamics.  

\textit{Ring FDSs---} Let us consider a cylindrical symmetric ring  FDS in a homogeneous system and the ring center is located at the origin.  The dynamics is cylindrically  symmetric and the corresponding  wavefunction ansatz in polar coordinates can be  parameterized by the ring radius  $R(t)$, i.e., $\psi_{\rm ring} (r,\varphi, t; V^2)=\psi_{\rm ring} (r,t;V^2)=\psi_{\rm ring} [r-R(t); V^2]$ with $V\rightarrow V(t)=dR(t)/dt$ being the ring radius velocity~\cite{SM}.  In 1D,  for $g_s/g_n=-1/2$ and $0< \tilde{q}<1$,  exact FDS solutions to Eqs.~\eqref{spin-1GPE} are available~\cite{FDSexact2022}. Based on the ring wavefunction ansatz and the exact solution in 1D,  the transverse magnetization and the mass superfluid densities of a ring FDS read 
\bea
\hspace{-10pt}
\label{magnetization}
F^{\rm I, II}_{\perp}(r,t)
&=&-\sqrt{ n_b^2-\frac{ q^2}{g_n^2}} \tanh \left\{\frac{r-R(t)}{\ell^{\rm I, II}[V^2(t)]}\right\}, \\ 
\hspace{-10pt}
n^{\rm I,II}(r,t) 
&=&n_b -\frac{g_n n_b-M V^2(t)\mp Q}{2 g_n}\sech ^2\left\{\frac{r-R(t)}{\ell^{\rm I, II}[V^2(t)]}\right\}, 
\eea
where $\ell^{\rm I,II}[V^2(t)]= \sqrt{2\hbar ^2/M \left[g_n n_b-M V^2(t)\mp Q\right]}$, 
$Q=\sqrt{M^2 V^4(t)+q^2-2 g_n M n_b V^2(t)}$, and the minus (plus) sign in front of $Q$ specifies type-I (II) FDS. At the speed limit $V=c_{\rm FDS}=\sqrt{g_nn_b/M}\sqrt{1-\sqrt{1-\tilde{q}^2}}$, $Q=0$ and the two types become identical.

\textit{EOM of the ring radius---} 
Let us firstly evaluate the ring FDS energy $\delta K_{\rm ring}$ and $\delta K_{\rm ring}=\int d^2\mathbf{r} \, ({\cal H}[\psi_{\rm ring}]-{\cal H}[\psi_g])+\mu \delta N$, where $\delta N=\int  d^2\mathbf{r}  \, (n_b-n_{\rm ring})$ is the depleted atom number,  
%$K_{\rm ring}= \int d^2\mathbf{r} \, ({\cal H}[\psi_{\rm ring}]-\mu n) $, $K_g= \int  d^2\mathbf{r}  \,({\cal H}[\psi_g]-\mu n_b)$, 
$\psi_g$ is the ground state wavefunction, and $\mu = (g_n + g_s)n_b + q/2$ is the chemical potential. 
Using the exact FDS wavefunctions~\cite{FDSexact2022} and the ring FDS wavefunction   ansatz $\psi_{\rm ring} [r-R(t)]$~\cite{SM}, we obtain the ring FDS  energy  
\bea
\hspace{-5mm}
&&\delta K_{\rm ring} [R(t),V^2]=\\
\hspace{-5mm}
&&\frac{4 \pi  \hbar ^4 \left\{\left[\exp\left(\frac{2 R(t)}{\ell^{\rm I,II}(V^2)}\right)+1\right]^2 \log \left[\exp\left(\frac{2 R(t)}{\ell^{\rm I,II}(V^2)}\right)+1\right]-\exp\left(\frac{2 R(t)}{\ell^{\rm I,II} (V^2)}\right)\right\}}{3 g_n M^2 \ell^{\rm I,II}(V^2)^2 \left[\exp\left(\frac{2 R(t)}{\ell^{\rm I,II}(V^2)}\right)+1\right]^2}. \nn
\eea
In the hydrodynamic limit,  i.e.,  $R(t) \gg \ell^{\rm I, II}(V^2)$, 
\bea
\delta K_{\rm ring} [R(t),V^2]=\frac{8 \pi  R(t) \hbar ^4}{3 g_n M^2 \ell^{\rm I,II}(V^2)^3}=2 \pi R(t) \delta K^{\rm I,II}(V^2)
\eea
where $\delta K^{\rm I,II}(V^2)=4 \hbar ^4/3 g_n M^2 \ell^{\rm I,II}(V^2)^3$ is the FDS energy in 1D systems and $\delta K^{\rm II}(V^2)>\delta K^{\rm I}(V^2)$~\cite{FDSexact2022}. 
The ring FDS energy conservation gives rise to
\bea
\left[\frac{R(t)} {R(0)}\right]^{1/3} = \frac{\ell^{\rm I,II}[V(t)]} {\ell^{\rm I,II}[V(0)]}, 
\label{EOM}
\eea
 which is equivalent to the following Newton equation describing dynamics of an elastic ring: 
\bea
\rho^{\rm I,II}_{\rm M} R \frac{d^2R}{dt^2}=-\sigma^{\rm I,II},
\label{tensionEOM}
\eea
where the ring tension $\sigma^{\rm I,II}(V^2)=\partial \delta K_{\rm ring}[R,V^2] /\partial (2 \pi R) = \delta K^{\rm I,II}(V^2)>0$ and the ring inertial mass density 
$\rho^{\rm I, II}_{\rm M}(V^2)=\pm (3 M/2) \sigma^{\rm I,II}(V^2)/Q(V^2)$ takes plus (minus) sign for type-I (II) FDSs and diverges at the speed limit ($Q=0$) when the type transition occurs.  Note that here the force acting on the ring FDS element, induced by the ring curvature, is different from the force acting on a FDS by the surrounding fluid in a 1D trapped system~\cite{FDSspincorrection}.  

We rewrite Eq.~\eqref{tensionEOM} as 
\bea
R\frac{d^2 R}{dt^2}= \frac{2}{3}\left[\kappa^{\rm I,II} R^{-2/3}+\left(\frac{dR}{dt}\right)^2 -\frac{g_n n_b}{M}\right],
\label{classicalring}
\eea
where $\kappa^{\rm I,II}=(9g^2_n/8 \pi^2 \hbar^2 M^2)^{1/3} (E^{\rm I,II}_0)^{2/3}$ and $E^{\rm I,II}_0$ is the initial ring FDS energy.  Let us denote the right-hand side of Eq.~\eqref{classicalring} as ${\cal T}$ and the condition ${\cal T}=0$ gives rise to the equilibrium position $R_c=(g_nn_b-M c^2_{\rm FDS}/\kappa^{\rm I,II})^{-3/2}$. It can be shown that for $R>R_c$,${\cal T}<0$ while for  $R<R_c$, ${\cal T}>0$ ~\cite{footnoteperiodic}. Hence Eq.~\eqref{classicalring} and therefore Eq.~\eqref{tensionEOM} must admit {\em period solutions}.   Comparing Eq.~\eqref{classicalring} to the inviscid Rayleigh-Plesset equation (RPE) ~\cite{brennen2014cavitation}, here $\kappa^{\rm I,II} R^{-2/3}=\tilde{\sigma}(R)/n_b R$ formally corresponds to the tenson term in the RPE however with radius-dependent tension $\tilde{\sigma}(R)=n_b\kappa^{\rm I,II} R^{1/3}$ and the term $\left(dR/dt\right)^2$ takes opposite sign.  These differences yield distinct dynamics. 

We emphasize that Eq.~\eqref{tensionEOM}  can be applied to general relevant ring solitons.  Taking an example of  a ring dark soliton in a uniform scalar BEC,  $\sigma^{\rm scalar}=(4 \hbar M/3g) (g n^s_b/M-V^2)^{3/2} $ and $\rho^{\rm scalar}_{\rm M}=-(4 \hbar M/g) (g n^s_b/M-V^2)^{1/2}$, where $g$ is the interaction strength and $n^s_b$ is the ground state superfluid density.  Applying Eq.~\eqref{tensionEOM},  we find that $d^2 R_{\rm scalar}/dt^2= [g n^s_b/M-(d R_{\rm scalar}/dt)^2]/3  R_{\rm scalar}$,  identical to the result reported in Refs.~\cite{ringdarksolitonKivshar, Ringdarksoliton2003}. 

\begin{figure}[htp] 
	\centering
	\includegraphics[trim = 0mm 0mm 0mm 0mm, clip, width=0.46\textwidth]{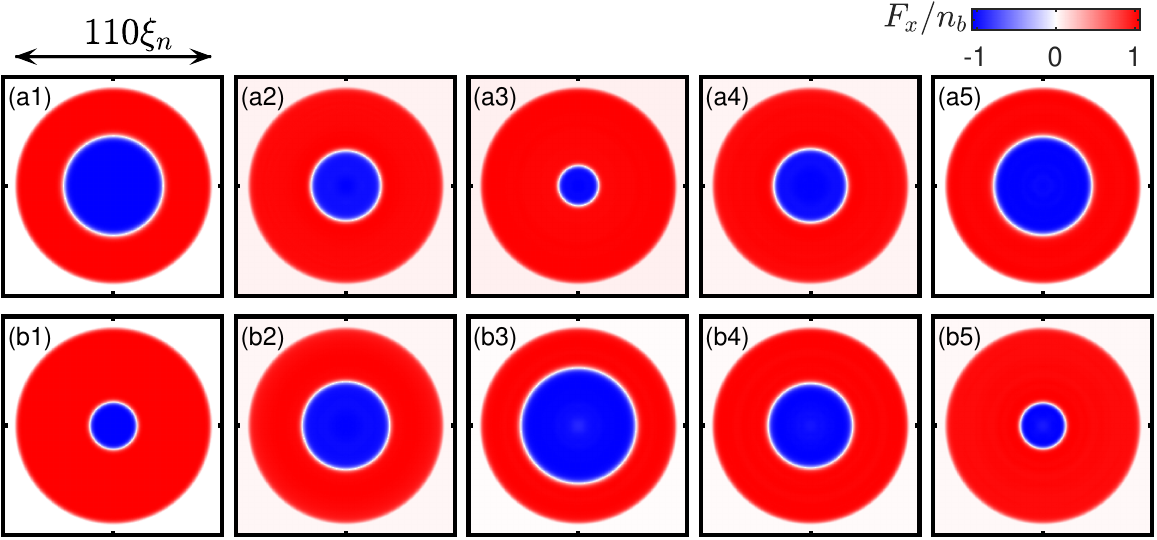}
	\vfill
	\hspace{-0.49cm}
	\includegraphics[trim = 0mm 0mm 0mm 0mm, clip,width=0.238\textwidth]{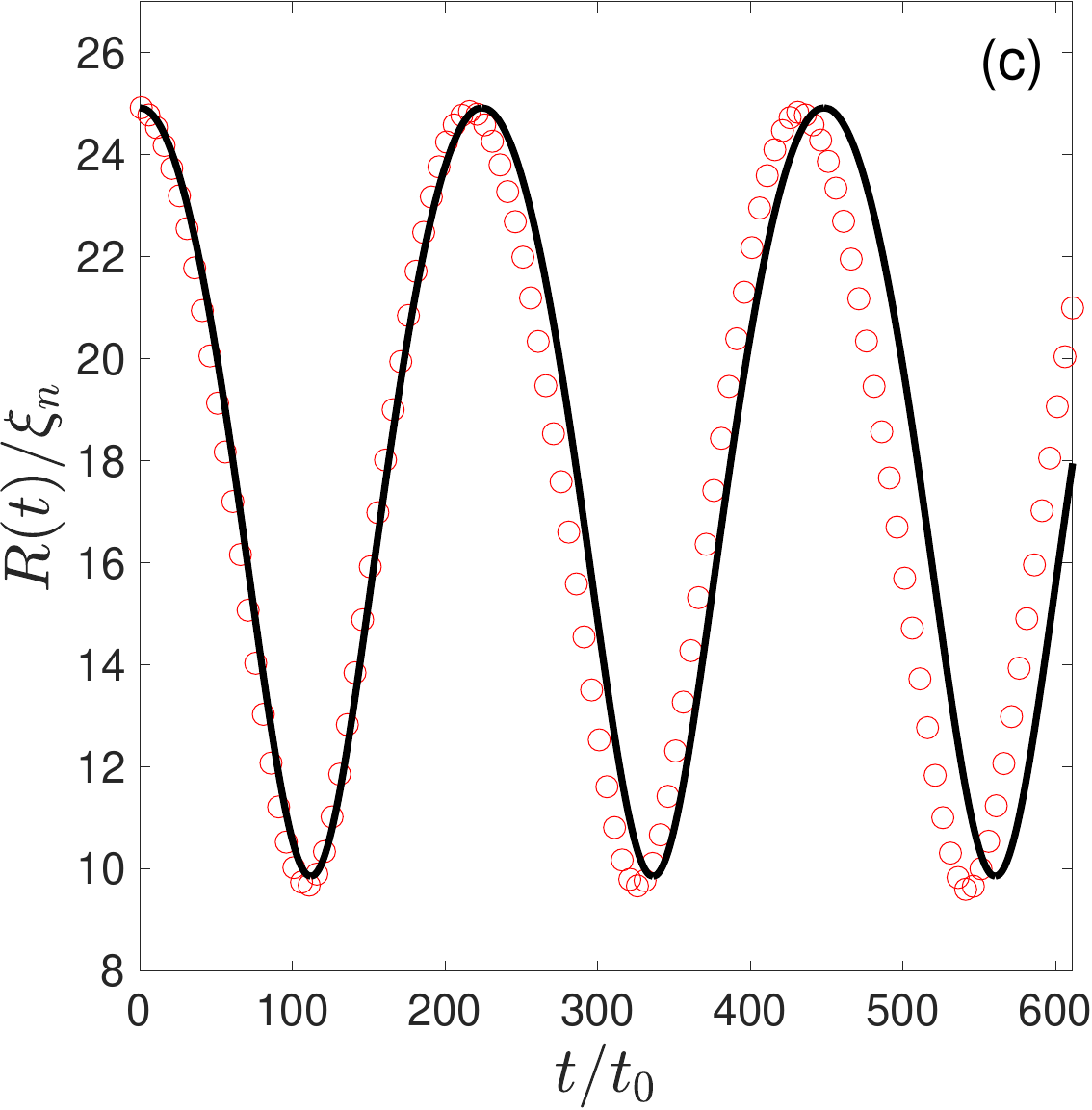}
	\hspace{0.08cm}
	\includegraphics[trim = 0mm 0mm 0mm 0mm, clip, width=0.243\textwidth]{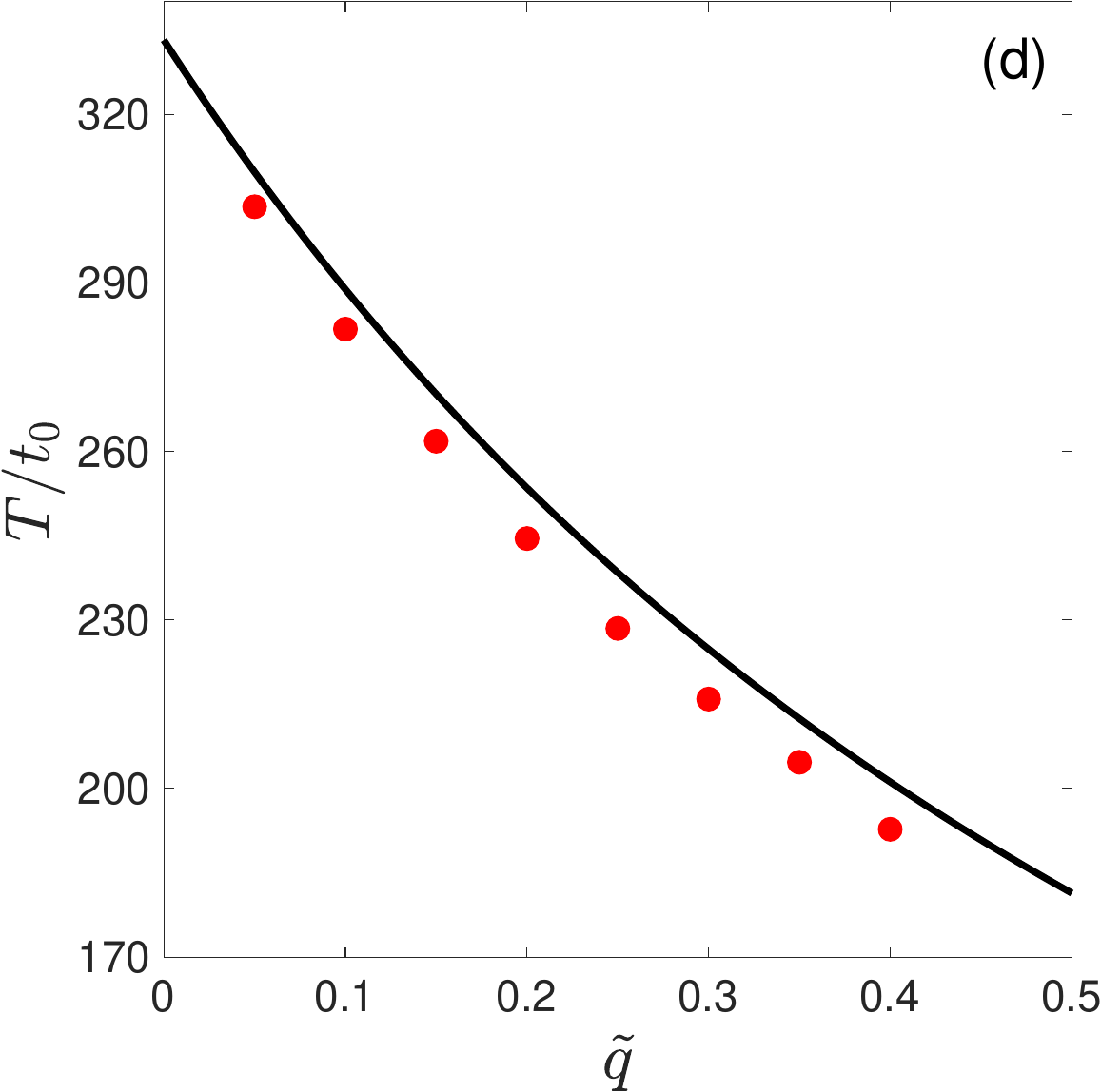}
	\caption{Oscillations of ring FDSs in a homogeneous  quasi-2D spin-1 BEC confined by a hard-wall potential, preserving the domain wall characteristics ($F_{\perp}=0$).  (a1-a5) and  (b1-b5): One period of evolution of a ring FDS initially prepared at type-I with $R_0=25\xi_n$ and type-II with $R_0=12 \xi_n$, respectively.  Here $g_s/g_n=-0.5$, and $\tilde{q}=0.3$. (c) Comparison between the analytical prediction [Eq.~\eqref{trajectory}] (solid lines) and numerical results (markers) for the evolution of the ring radius at $\tilde{q}=0.3$. (d) Comparison between the analytical prediction [Eq.~\eqref{oscillationperiod}] (solid lines) and numerical results (markers) of the ring oscillation period as a function of $\tilde{q}$ for $R_0=25 \xi_n$.} 
	\label{f:dynamicsring}
\end{figure}

%\bea
%\rho^{\rm I}_{\rm M}(V^2)&&=\frac{3 M}{2} \frac{T^{\rm I}(V^2)}{Q(V^2)} \\
%\rho^{\rm II}_{\rm M}(V^2)&&=-\frac{3 M}{2} \frac{T^{\rm II}(V^2)}{Q(V^2)} 
%\eea
\textit{Oscillations of the ring radius---} 
Let us prepare a type-I FDS ring with radius $R_0=R(0)$ and zero initial radius velocity [$V(t=0)=0$]. Its evolution in the first half period undergoes the following two stages: 
\bea
\hspace{-3mm}{\rm I}: \left[\frac{ R(t)}{R_0}\right]^{1/3} =\frac{\ell^{\rm I}[V^2(t)]}{\ell^{\rm I}[0]}; \quad {\rm II}: \left[\frac{R(t)} {R(t_c)}\right]^{1/3} = \frac{\ell^{\rm II}[V^2(t)]} {\ell^{\rm II}[c^2_{\rm FDS}]}.
\label{EOMharfperiod}
\eea
At stage I ($0<t<t_c$),  the type-I  FDS ring starts to shrink at an accelerating rate [Fig.\ref{f:dynamicsring}(a1)(a2)] and the radius velocity $V$ reaches the FDS speed limit $c_{\rm FDS}$ at $t=t_c$ where the type-I ring FDS converts to a type-II ring  FDS~\cite{FDSexact2022} and $R(t_c)=R_c=\chi_c R_0$ with $\chi_c=\left[(1-\tilde{q})/(1+\tilde{q})\right]^{3/4}$. At stage II, the ring (type-II) radius keeps shrinking while slows down due to the sign change of the ring inertial mass density. 
Later at $t=t_c$ the ring radius reaches its minimum value $R(t_s)=R_0 \chi_s$ with $V=0$, where $\chi_s=\left[(1-\tilde{q})/(1+\tilde{q})\right]^{3/2}$.  Clearly here $R(t_c)>R(t_s)$.  
The second half periodic motion ($t_c<t<t_s$) is the time-reversed evolution of the first half periodic motion and it begins with an expansion of the FDS ring (type-II) [Fig.\ref{f:dynamicsring}(a3)(a4)] with later returning the initial configuration [Fig.\ref{f:dynamicsring}(a5)].  
For the oscillation described above,  we find an exact solution to Eq.~\eqref{EOMharfperiod} within the time window $0<t<t_s$ [Fig.~\ref{f:dynamicsring}(c)]~\cite{SM}:
\bea
t(R)&=&\frac{3 \sqrt{M} R_0 \left\{(g_n n_b+q) g(R)f(R)-2 g_n  n_b \arctan \left[\frac{g(R)}{f(R)}\right]\right\}}{\sqrt{2} (g_n n_b+q)^{3/2}} \nn\\
&+&\frac{3 \pi \sqrt{M} g_n n_b R_0}{\sqrt{2} (g_nn_b+q)^{3/2}} ,
\label{trajectory}
\eea
where $f(R)=\sqrt{1-\left(R/R_0\right)^{2/3}}$ and $g(R)=\sqrt{\left(R/R_0\right)^{2/3}-\chi_s^{2/3}}$. 
Using Eq.~\eqref{trajectory},  we obtain that $t_s=t[R(t_s)]$. Hence the oscillation period of the ring FDS reads
\bea
T=2 t_s=\frac{3 \sqrt{2 M} \pi  g_n n_b}{ (g_n n_b+q)^{3/2}} R_0,
\label{oscillationperiod}
\eea
and the oscillation amplitude is $R_0 (1-\chi_s)$. The radius oscillation period decreases as decreasing $q$ [Fig.~\ref{f:dynamicsring}(d)].  When $q\rightarrow0$, $R(t)=R_0$,  the orbital  motion is frozen due to the magnetization conservation.  However in this limit 
$T\rightarrow T_0\equiv 3 \sqrt{2} \pi \sqrt{M/g_n n_b} R_0 \neq 0$, indicating remaining internal oscillations at the core which will be discussed in details below.  Note that for a type-II ring FDS  initially prepared,  the ring exhibits oscillations starting with an expansion [Fig.\ref{f:dynamicsring}(b1)-(b5)].

%For type-I FDSs, we obtain
%\bea
%-1=\frac{3 M}{2Q(V^2)}  r(t) \frac{d^2r(t)}{dt^2} \\
%\eea
%and for type-II FDSs we have 
%\bea
%1=\frac{3 M}{2Q(V^2)} r(t) \frac{d^2r(t)}{dt^2}
%\eea
%\begin{widetext}
%\bea
%t&&=\frac{3 \sqrt{M} r_0 \left((g_n n_b+q) \sqrt{\left(1-\left(\frac{r(t)}{r_0}\right)^{2/3}\right) \left(\left(\frac{r(t)}{r_0}\right)^{2/3}-\chi_c^{2/3}\right)}-2 g_n  n_b \arctan \left(\sqrt{\frac{\left(\frac{r(t)}{r_0}\right)^{2/3}-\chi_c^{2/3}}{1-\left(\frac{r(t)}{r_0}\right)^{2/3}}}\right)\right)}{\sqrt{2} (g_n n_b+q)^{3/2}}+\frac{3 \pi \sqrt{M} g_n n_b r_0}{\sqrt{2} (g_nn_b+q)^{3/2}} \nn\\
%\label{trajectory}
%\eea
%\end{widetext}

\textit{Parameterization---} In order to describe the  internal oscillations indicated by Eq.~\eqref{oscillationperiod}, we introduce a parameter $\Theta$ such that
\bea Q[V^2]&&=q \cos \Theta  \quad \text{for}  \quad 0<\Theta<\pi/2; \\
  Q[V^2]&&=-q \cos \Theta   \quad \text{for}  \quad \pi/2 <\Theta < \pi.
\eea 
Then $V^2=(g_n n_b/ M)\left(1-\sqrt{1-\tilde{q}^2 \sin ^2\Theta }\right)$ and $\ell(\Theta)= \sqrt{2 \hbar ^2/\left[M g_n n_b \left(\sqrt{1-\tilde{q}^2 \sin^2\Theta}-\tilde{q} \cos \Theta \right)\right]}$. Varying $\Theta$ monotonically, the FDS transfers  from type-I ($0<\Theta<\pi/2$) to type-II ($\pi/2<\Theta<\pi$) continuously  and hence the parameter $\Theta$ unifies the two types of FDSs~\cite{footnoteextension}.  When $\Theta=0$, $V^2=0$ and when $\Theta=\pi/2$, $V^2=c^2_{\rm FDS}$.  The exact wavefunction of a line (along the $y$-axis) FDS then reads: 
$\psi_{\pm 1}(\zeta,\Theta)=\sqrt{n^{\pm 1}_b}\left\{\alpha(\Theta) \tanh\left[\zeta/\ell(\Theta)\right]+i \delta (\Theta)\right\}$, and 
$\psi_0(\zeta,\Theta)=-\sqrt{n^{0}_b} \left\{\alpha(\Theta)+i \,\delta(\Theta) \tanh\left[\zeta/\ell(\Theta)\right]\right\}, $
where  $\zeta=x-V t$, 
$\alpha(\Theta)=-\sqrt{\left[1+ 1/\tilde{q}-2\hbar^2/ M q \ell^2(\Theta)  \right]/2}$,
$\delta(\Theta)=\sqrt{\left[1- 1/\tilde{q}+2\hbar^2/ M q \ell^2(\Theta) \right]/2}$, 
and $\alpha^2(\Theta)+\delta^2(\Theta)=1$.

%When $q\rightarrow 0$,
%\bea
%\alpha(\Theta)=-\sqrt{(1+\cos \Theta)/2} \\
%\delta(\Theta)=\sqrt{(1-\cos \Theta)/2}
%\eea

%\bea
%\psi^{\rm I}_{\pm 1}(x,t)= \sqrt{n^{\pm 1}_b}\left[\alpha^{\rm I} \tanh\left(\frac{r-R(t)}{\ell^{\rm I}}\right)+i \delta^{\rm I} \right] \\
%\psi^{\rm I}_0(x,t)=\sqrt{n^{0}_b} \left[-\alpha^{\rm I}-i \,\delta^{\rm I}  \tanh\left(\frac{x-Vt}{\ell^{\rm I}}\right)\right] \\
%\psi^{\rm II}_{\pm 1}=\sqrt{n^{\pm 1}_b}\left[-\alpha^{\rm II}-i \, \delta^{\rm II} \tanh\left(\frac{r-R(t)}{\ell^{\rm II}}\right)\right]\\ 
%\psi^{\rm II}_{0}=\sqrt{n^{0}_b}\left[\alpha^{\rm II} \tanh\left(\frac{r-R(t)}{\ell^{\rm II}}\right) +i \, \delta^{\rm II}\right]  \\  	
%\alpha^{\rm I}=-\sqrt{\frac{q+M V^2+Q}{2q}} ,\quad \delta^{\rm I}=\sqrt{\frac{ q-M V^2-Q}{2 q}}  \\
% \delta^{\rm II}=-\sqrt{\frac{q+ M V^2-Q}{2 q}}\\
% \alpha^{\rm II}=-\sqrt{\frac{q-M V^2+Q}{2q}} \\
%\beta^{\rm I}=-\alpha^{\rm I} \\
%\kappa^{\rm I} 
%		=-\delta^{\rm I}\\  
%			\beta^{\rm II}=-\alpha^{\rm II}\\
%		\kappa^{\rm II}=-\delta^{\rm II}
%\eea

\textit{Breathing nematic core--} 
At the core of a FDS ring, $\mathbf{F}[r=R(t)]=0$, we hence consider the nematic degrees of freedom which are characterized by the nematic tensor  $N_{ij}=\psi^{\dag} \hat{N}_{ij} \psi$, where $\hat{N}_{ij}=(S_iS_j+S_jS_i)/2$ with $i,j \in \{x,y,z\}$.  In Cartesian representation, the spinor field $\bm{\psi}={\cal C} \psi =e^{i\phi} (\mathbf{u} +i \mathbf{v})$, where ${\cal C}$ is a unitary matrix [Eq.~\eqref{cartesian}]~\cite{SM}. 
In terms of $\mathbf{u} $ and $\mathbf{v} $,  $\mathbf{F}=2 \mathbf{u}\times \mathbf{v}$ and $N_{ij}=\psi^{\dag} \hat{N}_{ij} \psi=\delta_{ij} n-(\mathbf{u} \otimes \mathbf{u}+\mathbf{v}\otimes \mathbf{v})$.  The  eigenvectors  of $N_{ij}$ are $\{\mathbf{u}, \mathbf{v}, \mathbf{F} \}$ and the corresponding  eigenvalues are $\lambda_{\mathbf{u}}=(n-|\alpha|)/2$, $\lambda_{\mathbf{v}}=(n+|\alpha|)/2$ and $\lambda_{\mathbf{F}} = n$, where $\alpha=\psi^2_0-2\psi_{+1}\psi_{-1}$
is the spin-singlet amplitude satisfying $|\alpha|=2|u|^2-n \geq 0$ and $|\mathbf{F}|^2+|\alpha|^2=n^2$.  At the core, $\mathbf{F}=0$,  hence $\lambda_{\mathbf{u}}=0$ and $\lambda_{\mathbf{v}}=\lambda_{\mathbf{F}}=n$. The wavefunction ansatz of the ring FDS $\psi_{\rm ring}=\psi [\zeta \rightarrow r- R(t),\Theta]$ gives rise to the nematic tensor at the core [$r=R(t)$]: 
\bea
\label{nematictensorcore}
N_{yy}&=&\frac{n_b}{2}-2n^{\pm1}_b \delta^2(\Theta); \quad N_{zz}=\frac{n_b}{2}-n^{0}_b \alpha^2(\Theta); \\ N_{yz}&=&-\sqrt{2n^{\pm1}_bn^{0}_b} \delta (\Theta) \alpha (\Theta); \quad  N_{xy}=N_{xz}=N_{xx}=0, \nn
\eea
and the mass superfluid density 
\bea
n[r=R(t)]=\frac{2 g_n n_b+q \cos \Theta-\sqrt{g_n^2 n_b^2-q^2 \sin ^2\Theta}}{2 g_n},
\label{densitycore}
\eea
where the dynamics of $\Theta$ is given by 
\bea
t(\Theta)&&=\frac{3 \sqrt{M} R_0 \left\{(g_n n_b+q) f(\Theta) g(\Theta)-2 g_n n_b \arctan\left[\frac{g(\Theta)}{f(\Theta)}\right]\right\}}{\sqrt{2} (g_n n_b+q)^{3/2}} \nn\\
&&+\frac{3 \pi  g_n \sqrt{M} n_b R_0}{\sqrt{2} (g_n n_b+q)^{3/2}}
\eea
with $f(\Theta)=\sqrt{1-(1-\tilde{q})/\left(\sqrt{1+\tilde{q}^2 \cos ^2\Theta -\tilde{q}^2}-\tilde{q} \cos \Theta \right)}$ and $g(\Theta)=\sqrt{1-f^2(\Theta)-\chi_s^{2/3}}$.

\begin{figure}[htp] 
	\centering
	\includegraphics[trim = 0mm 0mm 0mm 0mm, clip,width=0.234\textwidth]{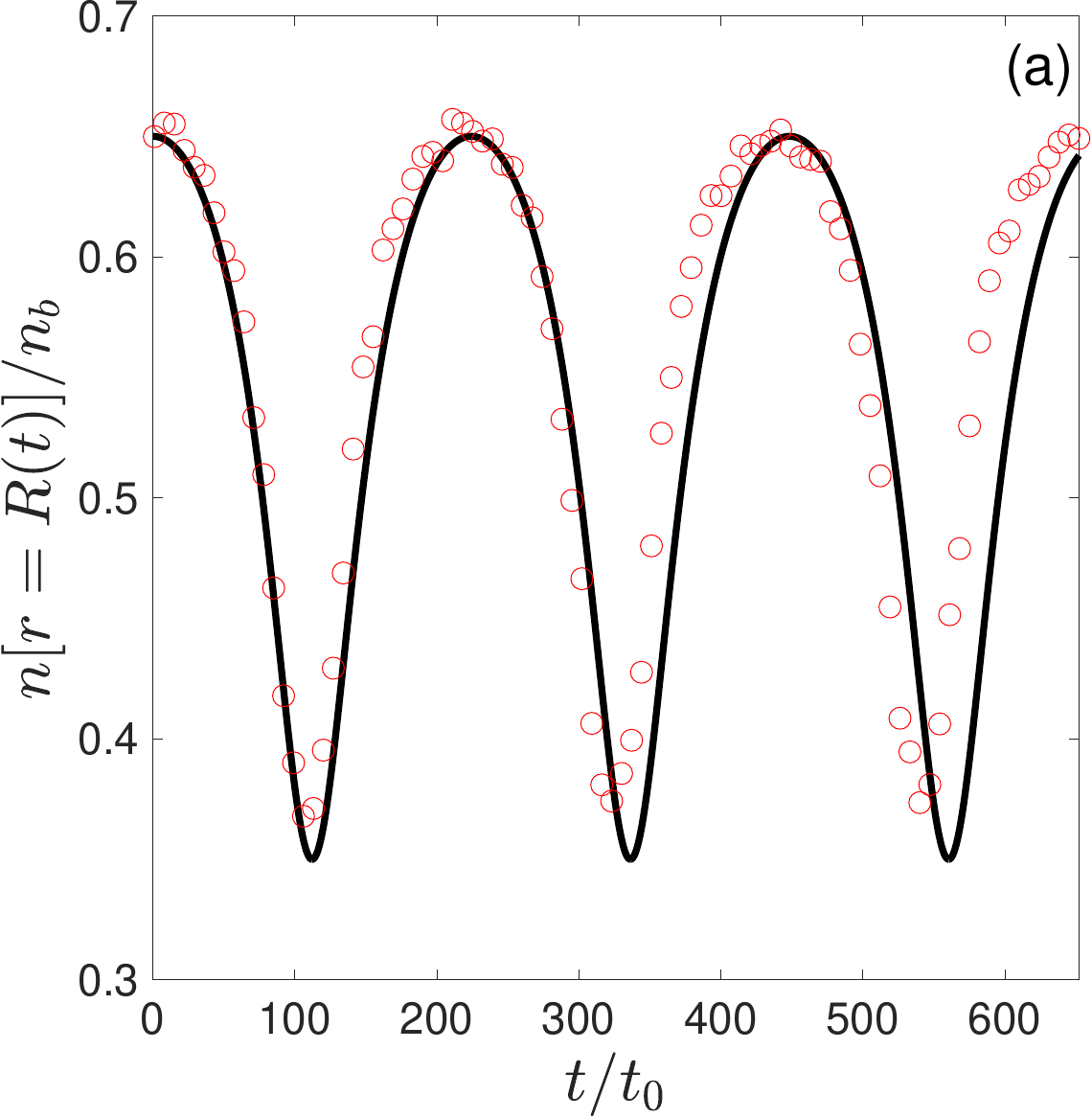}
	\hspace{0.08cm}
	\includegraphics[trim = 0mm 0mm 0mm 0mm, clip, width=0.234\textwidth]{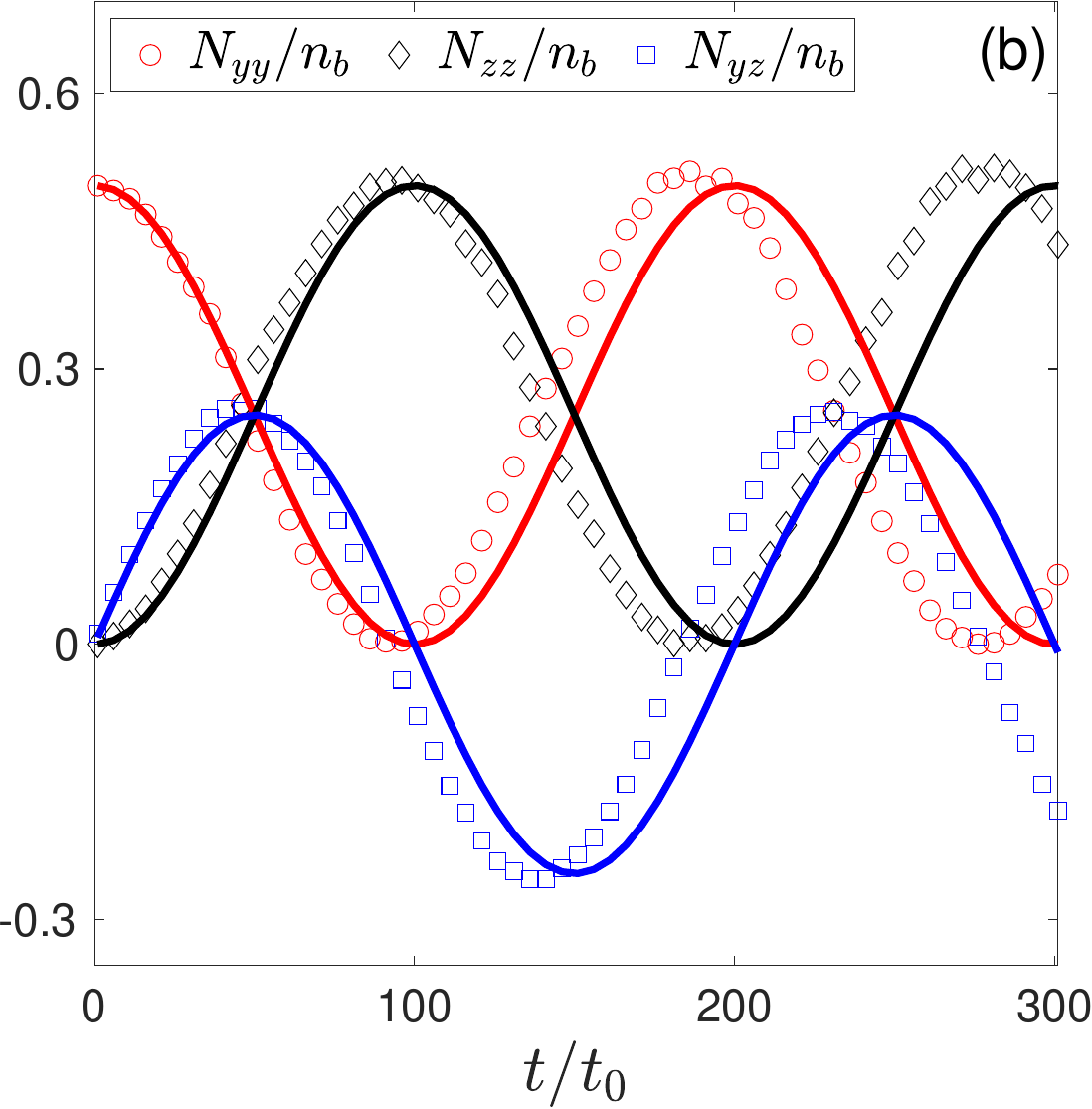}
	\caption{Dynamics of nematic degrees of freedom at the core of a ring FDS for $g_s/g_n=-0.5$. (a) Comparison between the analytical prediction [Eq.~\eqref{densitycore}] (solid lines) and the numerical results (markers) of the only non-zero eigenvalue of $N_{ij}$ (the mass superfluid density $n$) at the core of the oscillating ring FDS described in Fig.~\ref{f:dynamicsring} (a1-a5) at $\tilde{q}=0.3$. (b) Comparison between the analytical predictions [Eq.~\eqref{nematictensorcore}] (solid lines) and the numerical results (markers) of the dynamics of  the nemetic tensor components for $q=0$ and  $R_0=15\xi_n$. } 
	\label{f:dynamicsringnematic}
\end{figure}

 %At the core of the FDS ring, the wavefunction ansatz~Eq.~\eqref{wavefunctionansatz} gives rise to   
%\bea
%\vec{u}=\left(0,\sqrt{2 n^{\pm 1}_b} \delta (\Theta),-\sqrt{n^{0}_b} \alpha(\Theta) \right)
%\eea
For $q\neq0$,  the only non-zero eigenvalue of $N_{ij}$, i.e., the mass superfluid density at the core $n[r=R(t)]$, together with the ring radius, exhibits oscillations [Fig.~\ref{f:dynamicsringnematic}(a)]. This behavior is different from an oscillating FDS in trapped systems where the mass superfluid density at the core is nearly a constant~\cite{FDSexact2022, FDSspincorrection}. When $q\rightarrow 0$, the ring radius $R(t)$, the mass superfluid density $n$ and $\cal{N}\equiv\sqrt{\sum_{i,j} (N_{ij})^2}$ [$\textrm{SO}(3)$ spin-rotationally invariant]~\cite{FDScore2022} at the core become static  with $n[r=R(t)=R_0]\rightarrow n_b/2$  and ${\cal N}[r=R(t)] \rightarrow n_b/2$.  However the components of the nematic tensor $N_{ij}$ at the core still exhibit periodic motion [Fig.~\ref{f:dynamicsringnematic} (b)]~\cite{footnotenematicvector}. This is because, in the $q \rightarrow 0$ limit,  $\Theta$ is time-dependent and  
\bea
\Theta(t)=\left(\frac{3  \sqrt{M}  R_0}{\sqrt{2} (g_n n_b)^{1/2}}\right)^{-1} t=\frac{2\pi }{T_0} t,
\label{periodq0}
\eea
where we have used that $\alpha(\Theta)\rightarrow-\sqrt{(1+\cos \Theta)/2}$,  $\delta(\Theta)\rightarrow\sqrt{(1-\cos \Theta)/2}$,  $f(\Theta)\rightarrow 0$, $g(\Theta)\rightarrow 0$, and $\arctan\left[g(\Theta)/f(\Theta)\right] \rightarrow \pi/2-\Theta/2$. 
Clearly when $R_0 \rightarrow +\infty$ (a straight-line FDS), $\Theta(t) \rightarrow 0$, and it is the ring curvature that drives the motion of the nematic tensor components.   Consistently,  the oscillation period predicted by Eq.~\eqref{periodq0} coincides with Eq.~\eqref{oscillationperiod} in the $q\rightarrow 0$ limit. 
%It is the topological structure and the curvature of the ring FDS ensures the presence of  the breathing nematic core.  

At $q=0$, the system processes the $\textrm{SO}(3)$ spin-rotation symmetry.  For a spin-rotation $\psi \rightarrow {\cal  U}(\tau,\beta,\gamma) \psi$, where ${\cal  U}(\tau,\beta,\gamma)=e^{-i \tau s_z} e^{-i \beta s_y} e^{-i \gamma s_z}$ and $\tau$, $\beta$ and $\gamma$ are the Euler angles,  the nematic tensor changes  accordingly  $N_{ij} \rightarrow {\cal O}_{il} N_{lk} {\cal O}^{\rm T}_{kj}$, where ${\cal O}={\cal C}   {\cal  U} {\cal C} ^{\dag} $ is the fundenmental representation of $\textrm{SO}(3)$.  Hence the dynamics of the nematic tensor components at the core depends on the spin-rotation.  However it can be  shown  that the motion is intrinsic, i.e.,  no spin-rotation exists such that all the  components of the  nematic tensor become static~\cite{SM}.

\begin{figure}[htp] 
	\centering
	\includegraphics[width=0.41\textwidth]{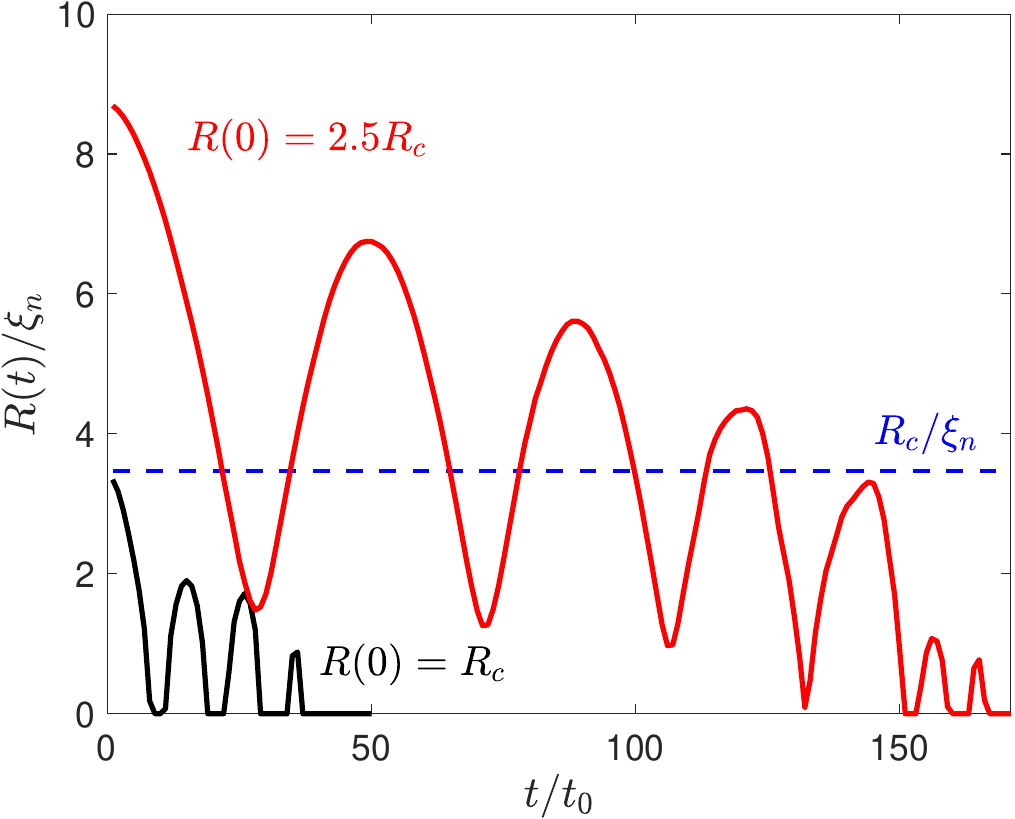}
	\caption{Evolution of the ring radius away from the hydrodynamic regime. Here $R(0)=2.5 R_c$ (red line) and $R(0)=R_c$ (black line).  The blue dashed line marks the  critical value $R_c$. Here $g_s/g_n=-0.5$ and $\tilde{q}=0.5$. As can be seen that a smaller ring has a shorter lifetime. } 
	\label{f:radiuscollapse}
\end{figure}

\textit{Dissipation, collapses and annihilation--} For a ring FDS away from the hydrodynamic regime,  emitting waves (including mass superflud density and magnetization density waves) becomes considerable. This causes the energy dissipation of the ring FDS and consequently the amplitude of  the oscillating ring gradually decreases (Fig.~\ref{f:radiuscollapse})~\cite{footnoteenergyloss}.  Let us consider a relatively small ring FDS initially prepared at type-I with zero radius velocity.  We find that there exists a critical value of the ring radius below which the first collapse ($R=0$) occurs followed by a few subsequent collapses and final annihilation (Fig.~\ref{f:radiuscollapse}).  Since only type-I FDSs annihilate~\cite{bai2025collisions}, the critical value of the ring radius can be estimated as following:  at $t \le t_c$ (still type-I ring FDS),  a collapse can occur if $R(t)\leq \ell^{\rm I}[V(t)^2]$, where $R(t)=R_0\left(\ell^{\rm I}[V^2(t)]/\ell^{\rm I}[0]\right)^3$.  This collapse condition can be rewritten as $R_0 \leq  \ell^{\rm I}[0]^3/\ell^{\rm I}[V(t)^2]^2 \leq R_c(q)$, where 
\bea
\hspace{-5mm} R_c(q) = \frac{\ell^{\rm I}[0]^3}{\ell^{\rm I}[c^2_{\rm FDS}]^2}=\sqrt{\frac{2 \hbar ^2}{M}} \frac{(g_n n_b+q)^{1/2}}{g_n n_b-q} 
\label{bound}
\eea
and we have used the fact that $\ell^{\rm I}[V^2(t)] \le \ell^{\rm I}[V(t_c)^2]= \ell^{\rm I}[c^2_{\rm FDS}]^2$ [$V(t)<V(t_c)=c_{\rm FDS}$]~\cite{FDSexact2022}.  Then $R_c(q)$ serves an estimate  of  the critical amplitude of the oscillating ring FDS below which the collapse happens. The bound $R_c(q)$ works well (Fig.~\ref{f:radiuscollapse}) and $R_c(q=0)=\ell^{\rm I}[0]$ as expected.

%\bea
%R(t_c)=r_0 \left(\frac{g_n  n_b-q}{g_n n_b+q}\right)^{3/4}.
%\eea

%Collapse may happen at different stages: happens at type-I or type-II 

%Happens at type-II: 
%\bea
%r(T_s)=\ell^{\rm II}[0]
%\eea
%
%
%\bea
%r_0  \left(\frac{g_n  n_b-q}{g_n n_b+q}\right)^{3/2}= \sqrt{\frac{2\hbar ^2}{M \left(g_n n_b+ q\right)}}
%\eea
%
%
%we find 
%\bea
%r_0=r_f(q)=\sqrt{\frac{2 \hbar^2}{M}} \frac{(g_n n_b+q)}{(g_n n_b-q)^{3/2}}
%\eea
%for $r_0<r_f(q)$, collision occurs. 
%
%These condition may work well for small $q$ but not for large $q$.  very tight bounds. 

%Collision occurs when 
%
%\bea
%\ell^{\rm I}[0] \leq r_0 \leq r_f(q)
%\eea

%For $\ell^{\rm I}[0] \leq r_0 \leq r_c(q)$,  collision happens for type-I FDSs and for $\ell^{\rm I}[0] < r_0 < r_c(q)$ collision happens at velocity $V< C_{\rm{FDS}}$.  For $r_0=r_c(q)$, collision occurs at the maximum velocity.  

%While for 
%\bea
% r_c(q) \leq r_0 \leq  r_f(q)
%\eea
%collision happens for  for type-II FDSs. For  $r_0 < r_f(q)$ collision happens at finite velocity and for  $r_0=r_f(q)$,  collision happens at zero velocity. 
%
%
%\bea
%\ell^{\rm I}[0]=\frac{\sqrt{2} \hbar }{\sqrt{M (g_nn_b-q)}}
%\eea
%

%\bea
%\hbar \omega  \left( {\begin{array}{cc}
%		u\\
%		v\\
%\end{array} } \right)=
%\left( {\begin{array}{cc}
%		{\cal L}_{\rm GP}+X-\mu & \Delta \\
%		-\Delta^{*} & -({\cal L}_{\rm GP}+X-\mu)^{*} \\
%\end{array} } \right)
%\left( {\begin{array}{cc}
%		u\\
%		v\\
%\end{array} } \right),
%\label{BdG1}
%\eea
\textit{Conclusion---}  
We study dynamics of ring ferrodark solitons in a homogeneous  ferromagnetic spin-1 Bose-Einstein condensate at zero temperature. The ring motion is that of a magnetic bubble.  We find that, in contrast to the usual expanding dynamics of ring dark solitons, the ring ferrodark soliton exhibits self-maintained nearly elastic oscillations with the dynamical core characterized by the periodic motion of the nematic tensor. When the ring radius is much larger than the soliton width, the equation of motion of the ring radius is derived and its exact solution is obtained. Excellent agreements are found between the analytical predictions and numerical simulations of spin-1  Gross-Pitaevskii equations.  In the absence of the magnetic field, the ring radius motion and the oscillation of the mass superfluid  density at the ring core freeze,  while the periodic motion of the nematic tensor components persists induced by the ring curvature.  
For small rings,  the ring oscillation becomes considerably dissipative, yielding ring collapses and annihilation at later times. In this work we focused on the exactly solvable parameter region while  the ring dynamics holds qualitatively for other values of $g_s/g_n$. This work explores the new regime of high dimensional topological soliton dynamics, i.e., ring soliton oscillations in a homogeneous system and this motion has not been encountered in the past.  This work will also motivate experimental investigations which are  within the scope of current ultracold-gas experiments~\cite{Dalibard2015,Gauthier16,Semeghini2018,Higbie2005,Huh2020a,MSexp1, MSexp2, prufer2022condensation}.

\textit{Acknowledgment---}
We thank P.~B.~Blakie,  H.Hu, X. Liu  for useful discussions. X.Y. acknowledges support from the National Natural Science Foundation of China (Grant No. 12175215, Grant No. 12475041), the National Key Research and Development Program of China (Grant No. 2022YFA 1405300) and  NSAF (Grant No. U2330401).  This paper is dedicated to Professor Ke Wu for his 80th Birthday.

%\bibliography{RingReferences}

\begin{thebibliography}{59}
\bibitem [{foo({\natexlab{b}})}]{footnoteextension}%
	\BibitemOpen
	\href@noop {} {\emph {\bibinfo {title} {\rm{Note that $\Theta$ can be
					extended to $[0,2\pi]$ in which case the FDS is type-I when
					$3\pi/2<\Theta<2\pi$ and is type-II when $\pi<\Theta<3\pi/2$}}}}\BibitemShut
	{NoStop}%
\bibitem [{foo({\natexlab{c}})}]{footnotenematicvector}%
	\BibitemOpen
	\href@noop {} {\emph {\bibinfo {title} {\rm{{At the core, $\mathbf{F}=0$,
						hence either $\mathbf{u} \parallel \mathbf{v}$ or $\mathbf{u}(\mathbf{v})=0$.
						It is worthwhile mentioning that at the core $\mathbf{v}\neq0$ hence the
						nematic degrees of freedom can not be captured only by the vector
						$\mathbf{u}$. For $q=0$, it can be shown that $|\mathbf{v}|^2$ (rotationally
						invariant) is not zero at the core (Fig.~\ref{f:nematicvector})~\cite{SM}. It
						is also clear that the nematic tensor $N_{ij}$ rather than $\mathbf{u}$ and
						$\mathbf{v}$ are better quantities to characterize the nematic degreees of
						freedom }}}}}\BibitemShut {NoStop}%
\end{thebibliography}


\pagebreak
\widetext
\begin{center}
	\textbf{\large Supplemental Material for `` Oscillating ring ferrodark solitons with breathing nematic cores in a homogeneous spinor superfluid ''}
\end{center}

\setcounter{equation}{0}
\setcounter{figure}{0}
\setcounter{table}{0}
\setcounter{page}{1}
\makeatletter
\renewcommand{\theequation}{S\arabic{equation}}
\renewcommand{\thefigure}{S\arabic{figure}}
\renewcommand{\bibnumfmt}[1]{[S#1]}
\renewcommand{\citenumfont}[1]{S#1}

This Supplemental Material includes the wavefunction ansatz of the ring FDS , the dynamics of nematic vectors at the core,  and other  necessary materials for supporting the main results presented in the main manuscript.

\section{Ring FDS wavefunctions}
The wavefunction ansatz of the ring FDS reads
\bea
\psi^{\rm I}_{\pm 1}(r,t)&&= \sqrt{n^{\pm 1}_b}\left\{\alpha^{\rm I} \tanh\left[\frac{r-R(t)}{\ell^{\rm I}}\right]+i \delta^{\rm I} \right\}, \nn\\
\psi^{\rm I}_0(r,t)&&=\sqrt{n^{0}_b} \left\{-\alpha^{\rm I}-i \,\delta^{\rm I}  \tanh\left[\frac{x-R(t)}{\ell^{\rm I}}\right]\right\} ,\nn\\
\psi^{\rm II}_{\pm 1}(r,t)&&=\sqrt{n^{\pm 1}_b}\left\{-\alpha^{\rm II}-i \, \delta^{\rm II} \tanh\left[\frac{r-R(t)}{\ell^{\rm II}}\right]\right\},\nn\\ 
\psi^{\rm II}_{0}(r,t)&&=\sqrt{n^{0}_b}\left\{\alpha^{\rm II} \tanh\left[\frac{r-R(t)}{\ell^{\rm II}}\right] +i \, \delta^{\rm II}\right\},  
\label{ringwavefunction}
\eea
where 
\bea
\alpha^{\rm I}=-\sqrt{\frac{q+M V^2+Q}{2q}} ,\quad \delta^{\rm I}=\sqrt{\frac{ q-M V^2-Q}{2 q}} , \quad
\alpha^{\rm II}=-\sqrt{\frac{q-M V^2+Q}{2q}}, \quad \delta^{\rm II}=-\sqrt{\frac{q+ M V^2-Q}{2 q}}
\eea
satisfying 
\bea
\alpha^{\rm I} \delta^{\rm I}=-\frac{\sqrt{q^2-(MV^2+Q)^2}}{2q} \quad  \text{and}  \quad \alpha^{\rm II} \delta^{\rm II}=\frac{\sqrt{q^2-(MV^2-Q)^2}}{2q}.  
\eea

\section{Nematic tensor: rotationally invariant quantities}
The Cartesian representation of the spinor field 
 $\bm{\psi}={\cal C} \psi =e^{i\phi} (\mathbf{u} +i \mathbf{v})$,  
where the unitary matrix is
\bea
\cal {C} =\left( {\begin{array}{ccc}
		-1/\sqrt{2}& 0 & 1/\sqrt{2} \\
		-i/\sqrt{2} & 0 & -i/\sqrt{2} \\
		0 & 1 & 0 \\
\end{array} } \right).
\label{cartesian}
\eea

For a spin-rotation $\psi \rightarrow {\cal  U}(\tau,\beta,\gamma) \psi$, the nematic tensor changes  accordingly
\bea
\tilde{N}_{ij}={\cal O}_{il} N_{lk}  {\cal O}^{\rm T}_{kj}.
\eea
Since 
\bea
\sum_{ij} \tilde{N}^2_{ij}=\sum_{ij} \sum_{\ell kmn}  {\cal O}_{i\ell} N_{lk} {\cal O}_{kj}^{\rm T} {\cal O}_{im} N_{mn} {\cal O}_{nj}^{\rm T}=\sum_{\ell kmn} \sum_{ij} {\cal O}^{\rm T}_{mi} {\cal O}_{i \ell}  {\cal O}_{kj}^{\rm T}  {\cal O}_{jn}  N_{mn}  N_{\ell k}=\sum_{\ell kmn} \delta_{m\ell}   \delta_{kn}  N_{mn}  N_{\ell k}=\sum_{k\ell} N^2_{\ell k}, 
\eea
we conclude that ${\cal N}(t)\equiv\sqrt{\sum_{i,j} N^2_{ij}}$ is rotationally invariant.  At $q=0$,  the dynamics of ${\cal N}(t)$ is frozen and ${\cal N}(t)=n_b/2$. 

For the  wavefunction ansatz Eq.~\eqref{ringwavefunction},  the nematic tensor reads
\bea
\hspace{-5mm}N=\left( {\begin{array}{ccc}
		0 & 0  & 0\\
		0 & \frac{n_b}{2}-2n^{\pm1}_b \delta^2(\Theta)& -\sqrt{2n^{\pm1}_bn^{0}_b} \delta (\Theta) \alpha (\Theta)\\ 
		0 & -\sqrt{2n^{\pm1}_bn^{0}_b} \delta (\Theta) \alpha (\Theta) & \frac{n_b}{2}-n^{0}_b \alpha^2(\Theta)\\ 
\end{array} } \right).
\eea
At $q=0$, $\alpha(\Theta)=-\sqrt{(1+\cos \Theta)/2}, \, \delta(\Theta)=\sqrt{(1-\cos \Theta)/2}$, and hence 
\bea
\hspace{-5mm}N=\left( {\begin{array}{ccc}
		0 & 0  & 0\\
		0 & \frac{n_b}{4}(1+\cos \Theta)&\frac{n_b}{4} \sin \Theta \\ 
		0 & \frac{n_b}{4} \sin \Theta & \frac{n_b}{4}(1-\cos \Theta)\\ 
\end{array} } \right)
\eea
for $0<\Theta<\pi$.  We now show that, at $q=0$, the dynamics of $N_{ij}$ at the ring FDS core is intrinsic. 
We firstly assume that there exists a rotation ${\cal O} $ such that $d [{\cal O} N{\cal O} ^{\rm T}]/dt=0$.  Since the transformation ${\cal O}  [dN/dt] {\cal O} ^{\rm T}$ does not change the eigenvalues of  $dN/dt$,  this means that all the eigenvalues of  $dN/dt$ are zero.  However it is easy to find that the eigenvalues of $dN/dt$ are $0$ and $\pm 1$.  Hence such a rotation does not exist.

\section{Nematic vectors at the core of the ring FDS}

\begin{figure}[htp!] 
	\centering
	\includegraphics[width=0.435\textwidth]{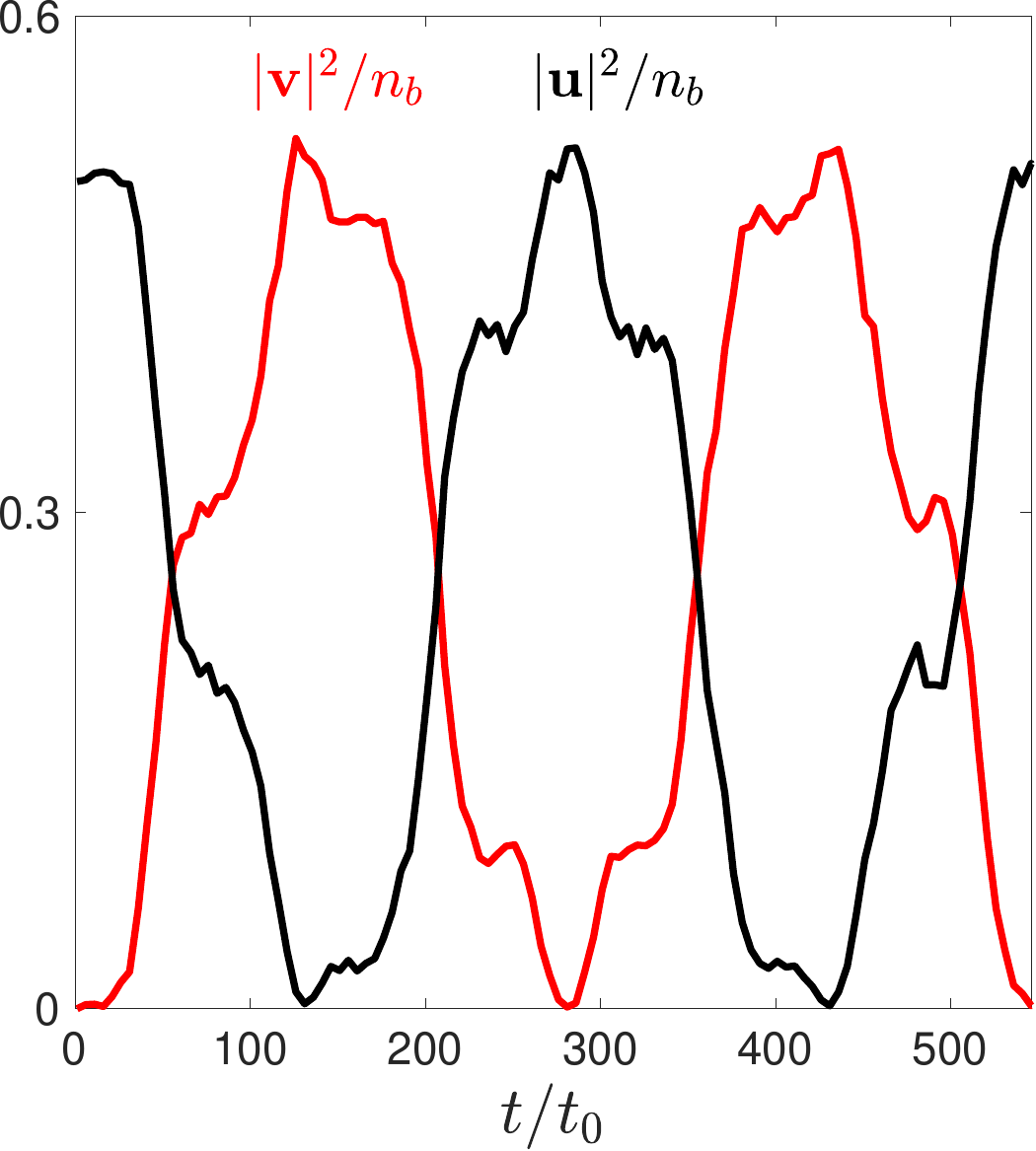}
	\caption{Evolution of $|\mathbf{u}|^2$ (back) and $|\mathbf{v}|^2$ (red) at the core of the ring FDS. Here $q=0$, $R_0=15\xi_n$ and $g_s/g_n=-0.5$, which are the same as parameters chosen in Fig.2(b) in the main text. } 
	\label{f:nematicvector}
\end{figure}

As shown in the main text,  at $q=0$ the wavefunction ansatz Eq.~\eqref{ringwavefunction} captures well the dynamics of the nematic tensor components at the ring FDS core.  However  the wavefunction ansatz does not describe the dynamics of nematic vectors $\mathbf{u}$ and $\mathbf{v}$ correctly.  
For instance, this wavefunction ansatz gives rise to that $\mathbf{v}=0$, however direct numerical simulations show that $\mathbf{v}\neq0$ (Fig~\ref{f:nematicvector}). At $q=0$,  the system processes the $\textrm{SO}(3)$ spin-rotation symmetry. For a spin-rotation $\psi \rightarrow {\cal  U}(\tau,\beta,\gamma) \psi$,  $\mathbf{u} \rightarrow {\cal O} \mathbf{u}$ and $\mathbf{v} \rightarrow {\cal O} \mathbf{v}$, where the orthogonal matrix ${\cal O}={\cal C}   {\cal  U} {\cal C} ^{\dag}$. 
Hence $|\mathbf{v}|^2$ is rotationally invariant and therefore if $\mathbf{v}\neq0$ for particular spin angles, it is not possible to find a spin rotation such that ${\cal O} \mathbf{v}=0$.

%Let us now consider a type-I FDS ring with radius $R$ and zero initial velocity ($V=0$) 
%
%
%
%
%\bea
%-\frac{1}{r(t)} dr(t)=\frac{3 M}{4Q(V^2)}   dV^2\\
%\eea
%
%\bea
%Q=\sqrt{M^2 V(t)^4+q^2-2 g_n M n_b V(t)^2}.
%\eea
%
%\bea
%-\log r(t)=-\frac{3}{4} \log \left(Q(V^2)-M V^2+g_n n_b\right)
%\eea
%
%
%\bea
%\log r(t)^2=\log \left(Q(V^2)-M V^2+g_n n_b\right)^\frac{3}{2} 
%\eea
%
%\bea
%c r(t)^2=\left(Q(V^2)-M V^2+g_n n_b\right)^\frac{3}{2} 
%\eea
%
%
%\bea
%c =\left(q+g_n n_b\right)^\frac{3}{2}/R^2
%\eea
%
%
%For $V=c_{\rm FDS}$, 
%
%\bea
%c r(t)^2=\left(g_n n_b \sqrt{1-\frac{q^2}{g_n^2 n_b^2}}\right)^\frac{3}{2} 
%\eea

%\bea
%\log r(t) =\frac{3}{2}  \tanh ^{-1}\left(\frac{q-Q(V^2)}{ M V^2}\right)+ C
%\eea
%
%\bea
%\tanh\left[\frac{2}{3} \log r(t) +C\right]= \frac{q-Q(V^2)}{ M V^2}
%\eea

\vspace{10mm}

\section{Two stages of the ring FDS evolution}
As described in the main text,   the evolution of the ring radius has two stages. At stage I, i.e.,  for $0<t<t_c$, the solution to Eq.~\eqref{EOMharfperiod} reads 
\bea
	t_{\rm I}(R)=t=\frac{3 \sqrt{M} R_0 \left((g_n n_b+q) \sqrt{\left[1-\left(\frac{R(t)}{R_0}\right)^{2/3}\right] \left[\left(\frac{R(t)}{R_0}\right)^{2/3}-\chi_c^{2/3}\right]}-2 g_n  n_b \arctan \left(\sqrt{\frac{\left(\frac{R(t)}{R_0}\right)^{2/3}-\chi_c^{2/3}}{1-\left(\frac{R(t)}{R_0}\right)^{2/3}}}\right)\right)}{\sqrt{2} (g_n n_b+q)^{3/2}}+\frac{3 \pi \sqrt{M} g_n n_b R_0}{\sqrt{2} (g_nn_b+q)^{3/2}} 
	\label{trajectory1}
\eea

where
\bea
\chi_c=\left(\frac{1-\tilde{q}}{1+\tilde{q}}\right)^{3/2}.
\eea
At stage II, i.e.,  for $t_c<t<t_s$, the solution to Eq.~\eqref{EOMharfperiod} reads 
\bea
	t_{\rm II}(R)=t=\frac{3 \left(\sqrt{g_n^2 n_b^2-q^2} \sqrt{\frac{2 g_n n_b \left[\frac{R(t)}{R(t_c)}\right]^{2/3}}{\sqrt{(g_n n_b-q) (g_n n_b+q)}}-\left[\frac{R(t)}{R(t_c)}\right]^{4/3}-1}+g_n n_b \arctan \left(\frac{\frac{2 g_n n_b}{\sqrt{(g_n n_b-q) (g_n n_b+q)}}-2 \left[\frac{R(t)}{R(t_c)}\right]^{2/3}}{2 \sqrt{\frac{2 g_n n_b \left[\frac{R(t)}{R(t_c)}\right]^{2/3}}{\sqrt{(g_n n_b-q) (g_n n_b+q)}}-\left[\frac{R(t)}{R(t_c)}\right]^{4/3}-1}}\right)\right)}{\sqrt{2} M \left[\frac{\sqrt{(g_n n_b-q) (g_n n_b+q)}}{M}\right]^{3/2}} R(t_c)
\eea
It is straightforward to  obtain that 
\bea
t_c=t_{\rm I}[R(t_c)] \quad \text{and}\quad  t_s=t_{\rm II}[R(t_s)].
\eea
Then the oscillation period reads
\bea
T=2 t_s=\frac{3 \sqrt{2 M} \pi  g_n n_b}{ (g_n n_b+q)^{3/2}} R_0.
\eea
We find that the two-stage evolution $t_{\rm I}(R)$ and  $t_{\rm II}(R)$ can be captured by a unified form, i.e., Eq.~\eqref{trajectory} in the main text within the time window $0<t<t_s$.

\end{document}